\documentclass{iau}

\usepackage{amsmath}
\usepackage{graphicx}
\usepackage{multirow}

\providecommand\apjl{The Astrophysical Journal Letters}
\providecommand\apj{The Astrophysical Journal}
\providecommand\Apj{The Astrophysical Journal}
\providecommand\jgr{Journal of Geophysical Research}
\providecommand\solphys{Solar Physics}

\begin{document}

\lefttitle{B.~T. Welsch \& C.~R. DeVore}
\righttitle{Reconnection in CME Structure \& Dynamics}

\jnlPage{1}{7}
\jnlDoiYr{2024}
\doival{10.1017/xxxxx}
\volno{388}
\pubYr{2024}
\journaltitle{Solar and Stellar Coronal Mass Ejections}
\aopheadtitle{Proceedings of the IAU Symposium}
\editors{N. Gopalswamy,  O. Malandraki, A. Vidotto \&  W. Manchester, eds.}

\title{The Role of Magnetic Reconnection in 
the Structure \& Dynamics of Fast Coronal Mass Ejections}

\author{Brian T. Welsch$^1$ \& C.~R. DeVore$^2$}

\affiliation{$^1$Natural \& Applied Sciences, University of Wisconsin - Green Bay, \\ 2420 Nicolet Drive, Green Bay, WI 54311, USA}

\affiliation{$^2$Heliophysics Science Division, NASA Goddard Space Flight Center, \\ 8800 Greenbelt Road, Greenbelt, MD 20771, USA}

\begin{abstract}
Both observations and models of flare-associated coronal mass ejections (CMEs) suggest that magnetic reconnection in an ejection's wake substantially increases the net, outward Lorentz force accelerating the CME.  A stronger outward force can cause a feedback loop, driving further magnetic reconnection in a ``reconnective instability.'' The flux accretion model captures this by relating reconnected flux, $\Delta \Phi_{\rm rec}$, and magnetic field strength, $B_{\rm CME}$, to increased outward Lorentz force, $\Delta F_r$. To better understand reconnection's role in CME dynamics, we analyze two snapshots from a 2.5D, MHD simulation of a breakout eruption.  Outward Lorentz forces increase substantially as reconnection proceeds, caused primarily by ``flank currents,'' which flow just inside the boundary of the rising ejection’s wake and parallel to its axis.  This model's reconnection jet also alters the ejection's internal structure, an effect that could be sought in observations.  Analyzing reconnection-induced Lorentz forces in 3D simulations could provide additional insights into CME dynamics.
\end{abstract}


\maketitle

\section{Introduction}
\label{sec:intro}

Magnetic reconnection is a rich topic (e.g., \citealt{Priest2000})
that is relevant to magnetic energy release in plasmas of the Sun, 
other stars, and many other physical systems.  In the case of solar
flares, reconnection is widely believed to be the central process
facilitating the release of stored magnetic energy (e.g.,
\citealt{McKenzie2002}).  For a flare-associated coronal mass ejection
(CME), {\em flare reconnection} \citep{Antiochos1999a} is believed to
occur in the wake of the rising ejection.  \citet{Antiochos1999a}
contrast flare reconnection with {\em breakout reconnection}, which
occurs in front of the erupting structure, and which they posit
initiates the eruption by removing overlying, restraining
flux. \citet{Lin2000} analyzed a 2.5D, Cartesian model of a CME, and
describe a feedback loop between flare reconnection and the upward
force accelerating the CME.  \citet{Moore2001} proposed that the
eruption is initiated by {\em tether cutting} reconnection under the
pre-eruptive structure, which reduces the effect of downward magnetic
tension from fields overlying that structure.  Once an eruption is
underway, tether cutting reconnection and flare reconnection are
indistinguishable.

Observationally, significant correlations have been found between
CMEs' speeds, v$_{\rm CME}$, and the amounts of magnetic flux
that reconnected in their associated flares, $\Phi_{\rm rec}$,
deduced from flare ribbons \cite{Qiu2005} or post-eruption arcades
\cite{Gopalswamy2017a}.  These correlations suggest, consistent with
the theoretical models above, that reconnection plays an important
role in altering the forces on ejections.  From the temporal
correlation between soft X-ray emission, interpreted as a proxy of
magnetic reconnection, and CME acceleration, \cite{Zhang2006} also
suggested that reconnection and CME acceleration operate in a feedback
loop early in the eruption process: more reconnection drives more
acceleration, which in turn drives further reconnection, and further
subsequent acceleration.

%Two popular models of CME initiation, the
%breakout model \cite{Antiochos1999a} and the tether-cutting model
%\cite{Moore2001} invoke reconnection to eliminate the inward Lorentz
%force from magnetic tension in overlying ``strapping fields'' that
%prevent an eruption from occurring.

%[In both cases, say something about the geometry, and ``flare
%  reconnection'' in the current sheet behind the ejection.]

Based upon these and other observations and models, \citet{Welsch2018}
presented the {\em flux accretion} model of eruptions, in which flare
reconnection below a rising ejection adds magnetic flux to the
ejection.  This (i) leads to momentum transfer from the outflow jet to
the ejection and (ii) increases the net, outward component of the
Lorentz force,$(\mathbf{J} \times \mathbf{B})/c$, acting on the
.ejection.  In the model's treatment of Lorentz forces, the accreted
underlying flux both reduces the net force due to downward magnetic
tension from overlying fields (the mechanism of tether cutting) and
increases the outward hoop force (e.g., \citealt{Chen1996}) from these
underlying fields exerting greater outward magnetic pressure on the
ejection. For both components of the total force, the flux accretion
model predicts that the change in outward force, $\Delta F_r$, due to the
reconnection of an amount of flux $\Delta \Phi_{\rm rec}$ scales as
\begin{equation}
  \Delta F_r \propto \Delta \Phi_{\rm rec} B_{\rm CME} ~, \label{eqn:f_scaling}
\end{equation}
where $B_{\rm CME}$ is a characteristic field strength in the vicinity
of the CME.  

In this brief paper, we explore how reconnection affects the Lorentz
force on models of CMEs. In Section \ref{sec:qual}, we consider a
highly idealized model of magnetic fields and currents in a CME source
region, and analyze the effect of reconnection upon these fields and
currents. Then, in Section \ref{sec:quant}, we quantitatively
investigate evolution of the Lorentz forces --- and the magnetic
fields and currents that contribute to it --- in two snapshots from a
numerical simulation of CME as flare reconnection proceeds.

\section{Qualitative Analysis of Reconnection's Effect on Lorentz Forces}
\label{sec:qual}

%Given that current, $\mathbf{J}$, plays in the Lorentz force,

\begin{figure}[t]
  \centerline{\vbox to 6pc{\hbox to 10pc{}}}
  \includegraphics[scale=.35]{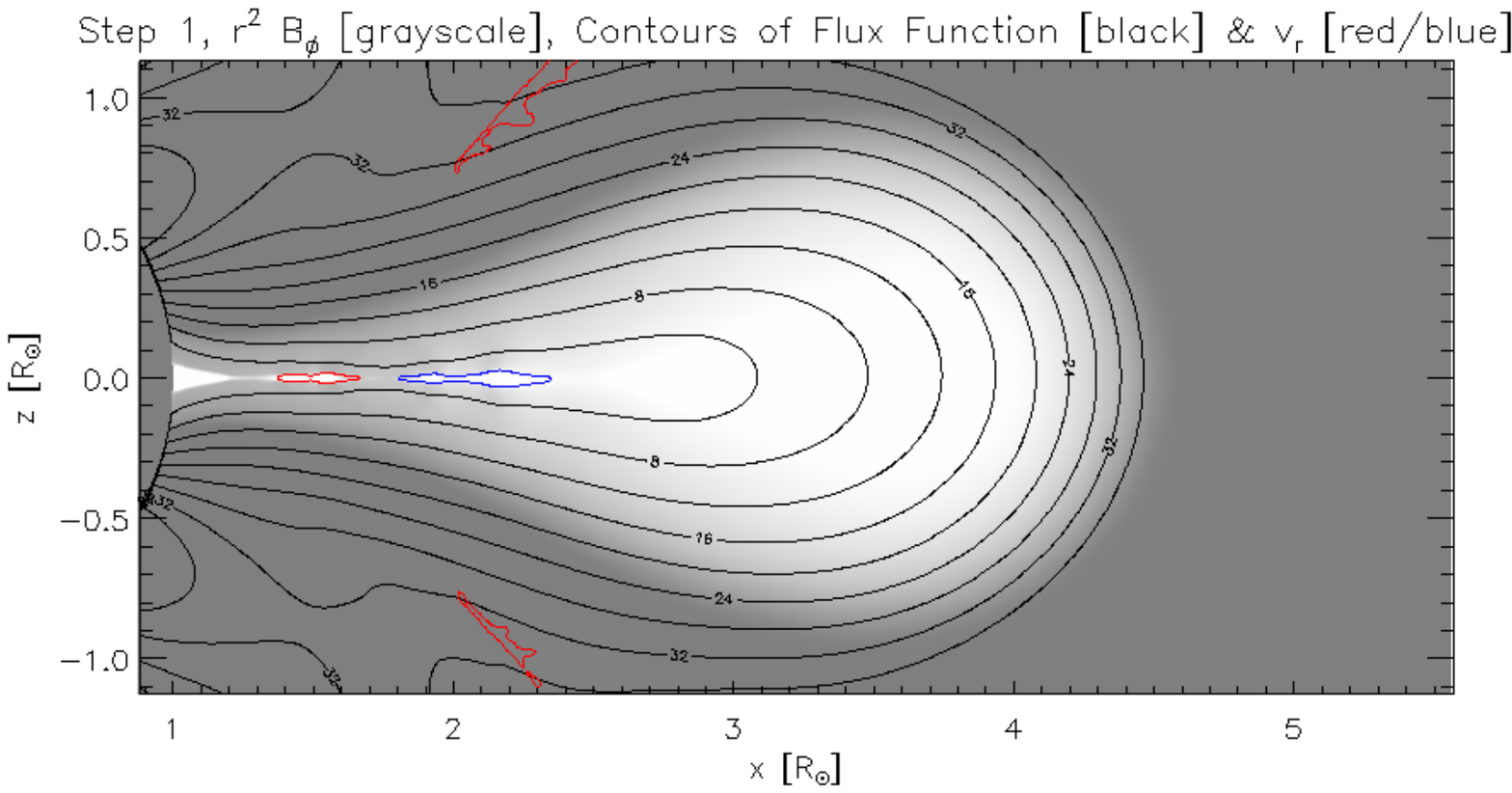}
  \caption{In this snapshot from the spherical, axisymmetric, 2.5D MHD
    simulation that we analyze, the simulation's spherical inner
    boundary ($r = R_\odot$) can be seen at frame left, $x$ increases to the
    right, $\theta$ increases clockwise from the Sun's north pole (out
    of the frame, at $z$ = +1) toward its equator (at $z$ = 0), and
    $\phi$ increases into the page.  The grayscale shows $B_\phi$,
    weighted by $r^2$ to keep the structure of $B_\phi$ visible as it
    weakens with increasing $r$.  The red and blue contours correspond
    to 100 km/s inward and outward radial velocities, respectively.
    The black lines show contours of the flux function, $f$,
    corresponding to poloidal projections of field lines.}
  \label{fig:geometry}
\end{figure}

We first consider the overall structure of the magnetic field and
currents in a generic eruptive configuration, and then consider how
these are relevant for flare reconnection and evolving Lorentz forces
during the eruption.  To illustrate the large-scale morphology of the
system, Figure \ref{fig:geometry} shows a snapshot from the
axisymmetric, 2.5D MHD simulation that we will analyze in the next
section.  (Details of the simulation are described there.)
The simulation employed spherical coordinates, and we interpolated its
physical values onto a uniform, Cartesian, (1024 $\times$ 1024) grid,
with $x \in [0, 6] R_\odot$ and $z \in [-3,3] R_\odot$.
The simulation's inner boundary, at $r = R_\odot$, corresponds to the base
of the corona and can be seen near the image's left edge.
%, with $x$ increasing to the right and $z$ increasing toward page top.
Co-latitude $\theta$ increases clockwise
from the Sun's north pole (out of the frame, at $z$ = +1) toward its
equator (at $z$ = 0), and $\phi$ increases into the page. 
The grayscale shows $B_\phi$, weighted by $r^2$ so its structure
remains visible even as it weakens with larger $r$.  The black lines
show contours of the flux function, $f$, with the poloidal field,
$\mathbf{B}_{\rm pol}$, obeying
\begin{equation}
  \mathbf{B}_{\rm pol} = \nabla f \times \nabla \phi ~,
%  ~~\rmbox{and}~~ A_\phi = f/(r \sin \theta)
  \label{eqn:flux_func}
\end{equation}
meaning contours of $f$ correspond to field lines of $\mathbf{B}_{\rm pol}$.
For $x \in [1.25,2.5] R_\odot$, this flux system's radial field is
outward below the $z = 0$ line and inward above it.  An intense
azimuthal current --- a large-scale current sheet --- flows between
these oppositely directed radial fields, and is directed out of the
page. In Figure \ref{fig:geometry}, this current is concentrated
between the innermost flux function contours straddling $z = 0$.
%Reconnection occurs on smaller-scale, more intense current sheets
%within this large-scale sheet.
The red and blue contours bracketing $(x,z) = (1.7, 0)$
correspond to 100 km/s inward and outward radial outflows,
respectively, from the flare reconnection region.

An eruptive flare invariably occurs above a photospheric polarity
inversion line (PIL), across which radial photospheric magnetic field
changes sign.  The horizontal photospheric field along a flaring PIL
almost always exhibits a significant {\em shear} component, i.e., a
component of $\mathbf{B}$ parallel to the PIL.  Associations between
magnetic shear and flare occurrence were noted long ago (e.g.,
\citealt{Hagyard1984}).  In addition to being flare-prone, sheared
fields along PILs have also been associated with increased likelihood
of CME occurrence (e.g., \citealt{Falconer2001, Falconer2003}).
Typically, sheared fields are localized in the vicinity of PILs, with
horizontal photospheric fields farther from PILs exhibiting less shear
(e.g., \citealt{Falconer2001}).  This decrease in shear with distance
from the PIL is also typically observed in overlying fields, inferred
from the directions of chromospheric H-$\alpha$ fibrils and coronal
loops' axes.
%
%The left panel of Figure \ref{fig:shear} illustrates shows an overhead
%view of this
%
%\begin{figure}[t]
%  \centerline{\vbox to 6pc{\hbox to 10pc{}}}
%  \includegraphics[scale=.4]{reconn_currents_3d_v2}
%  \caption{Field lines!
%  \label{fig:shear}
%\end{figure}
%

\begin{figure}[t]
  \centerline{\vbox to 6pc{\hbox to 10pc{}}}
  \includegraphics[scale=.5]{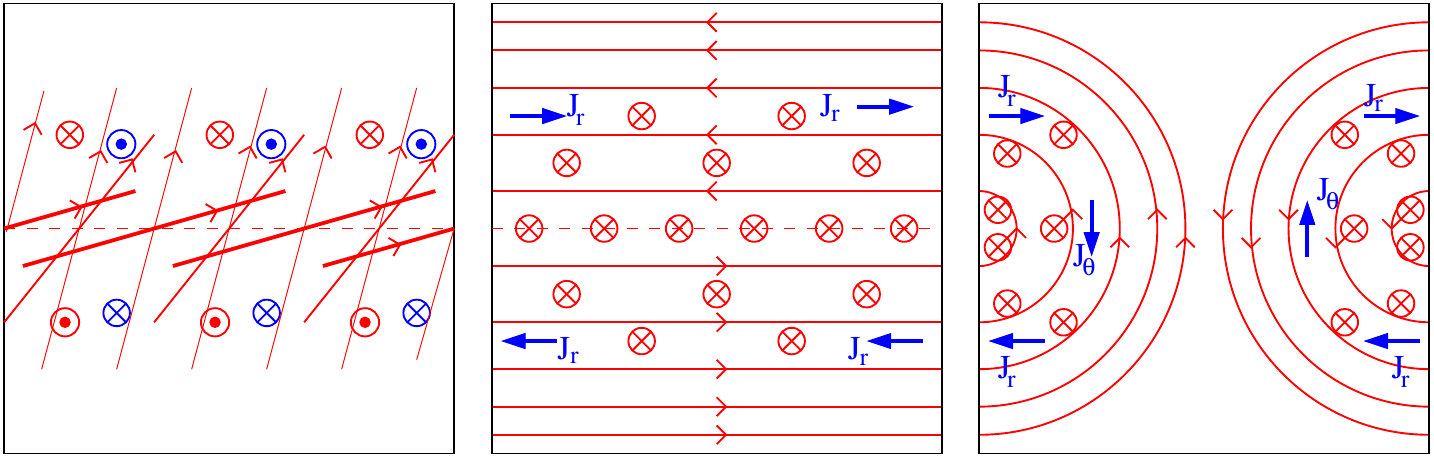}
  \caption{Left: An overhead view of magnetic structure near a sheared
    photospheric PIL (red dashed line). Solid red lines with arrows
    show magnetic fields above the photosphere, with thicker to
    thinner corresponding to fields at increasing altitudes. Red
    $\odot$ and $\otimes$ show polarity of the photospheric radial
    magnetic field. Blue $\odot$ and $\otimes$ symbols show the sheath
    current that isolates (or ``switches off'') the magnetic field's
    shear component (which points to the right).
    Middle: This side view, above and along the PIL from the left panel,
    shows flare reconnection, in which oppositely directed magnetic
    field components in the corona (solid red lines) reconnect across
    a current sheet (dashed red line, with its current directed out
    of the page) in the presence of guide field (red $\otimes$
    symbols).  Sheath currents associated with the localization of the
    guide field are also present (blue vectors).
    Right: Reconnection joins formerly oppositely directed fields into
    arched fields (solid red lines), and in the process reroutes the
    sheath currents (blue vectors), introducing oppositely directed
    currents between the post-reconnection flux systems. Each of these
    field and current components can, in principle, contribute to the
    radial Lorentz force that accelerates the ejection.}
  \label{fig:shear}
\end{figure}

Because the coronal plasma in CME source regions is magnetically
dominated (the plasma's $\beta$, its ratio of gas to magnetic
pressures, is much less than unity), the $(\mathbf{J} \times
\mathbf{B})/c$ (Lorentz) force is very small in slowly evolving,
pre-eruptive fields, i.e., such fields are very nearly {\em
  force-free.}  Thus, (i) $\mathbf{J} \ne 0$ implies that $\mathbf{J}$
must flow very nearly parallel to $\mathbf{B}$, and (ii) the presence
of a significant shear component of $\mathbf{B}$ implies a shear
component of $\mathbf{J}$.
The localization of the shear component of $\mathbf{B}$ in the
vicinity of the PIL implies that there must also be a current sheath
surrounding the sheared field, to isolate or ``switch off'' the shear component
of $\mathbf{B}$.
These elements are illustrated in the left panel Figure
\ref{fig:shear}, which shows an overhead view of magnetic structure
near a sheared photospheric PIL (red dashed line).  The red lines with
arrows show magnetic fields above the photosphere, with lines
progressing from thicker to thinner with increasing height. The red
$\odot$ and $\otimes$ show the polarity of the photospheric radial
magnetic field. The blue $\odot$ and $\otimes$ symbols show the sheath
current (similar to that in a solenoid) that isolates the magnetic
field's shear component (which points to the right).

Flare reconnection alters the connectivity of both preflare coronal
magnetic fields and current structures.  We are primarily interested
in how reconnection alters the radial component of the Lorentz force,
$F_r$,
\begin{equation}
c F_r = J_\theta B_\phi - J_\phi B_\theta 
  ~. \label{eqn:lorentz_r}
\end{equation}
Current in the reconnecting current sheet (which we assume is thin,
but not infinitely so) also flows parallel and above to the PIL; in
our geometry, this is $J_\phi$.  The shear component of
$\mathbf{B}$ corresponds to a guide field, i.e., a nonzero field
component along the reconnecting current sheet (see, e.g.,
\citealt{Priest2000}); in our geometry, this is $B_\phi$.
Reconnection in the presence of a guide field can introduce twist into
originally untwisted fields (e.g., \citealt{Wright1989,
  Chae1999}). The resulting fields can be highly twisted (e.g.,
\citealt{Longcope2007}) and generally carry currents.
Aspects of the reconfiguration of coronal fields and current
components are illustrated schematically in the middle and right
panels of Figure \ref{fig:shear}.  These panels depict a view in the corona 
along the guide field (i.e., above and along the PIL in the Figure's
left panel), showing the reversed field components that reconnect
(solid red lines), the guide field (red $\otimes$ symbols), and the
sheath currents (blue vectors) associated with the localization of the
shear component of $\mathbf{B}$.  In the middle panel, a current sheet
exists along the red dashed line, with current flowing out of the
page.  Reconnection joins oppositely directed fields (solid red lines
in middle panel) into arched fields (solid red lines in right panel),
which possess a $B_\theta$ component in our geometry.  The process
also reroutes the sheath currents (blue vectors), introducing
oppositely directed currents between the post-reconnection flux
systems, corresponding to a $J_\theta$ component.  The repulsion of
these antiparallel $J_\theta$ currents works against the attraction of
the parallel $J_\phi$ currents in the post-reconnection fields (both out of the page here).

In the following section, we investigate the evolution of $F_r$ as
flare reconnection proceeds in the model CME, and compare the
contributions from the $(J_\theta B_\phi)$ and $(J_\phi B_\theta)$ terms  
in Equation (\ref{eqn:lorentz_r}).

\section{Quantitative Analysis of Reconnection's Effect on Lorentz Forces}
\label{sec:quant}

%0241354         
%0254006 
%0266777

%We now investigate the behavior of Lorentz forces in an MHD simulation
%of CME as reconnection modifies the erupting regions magnetic fields
%and currents.
In this section, we analyze two snapshots from a 2.5D, axisymmetric,
breakout simulation using the Adaptively Refined Magnetohydrodynamic
Solver (ARMS)
%, performed by C.~.R. DeVore (private communication)
that
closely matches the simulations performed and analyzed by
\citet{Karpen2012} (hereafter ``KAD12'').  The run analyzed here
differs in only one potentially relevant detail: in the simulations
that we analyze, six levels of grid refinement were permitted, whereas
only five levels were enabled in KAD12.  The first snapshot, our Step
1 (the code's step 241354), corresponded to the onset of flare
reconnection, and at the second snapshot, our Step 2 (the code's step
254006), a substantial amount of flux has reconnected.
(Data from additional time steps could be analyzed, but computation of
the flux function for this simulation must be done {\em post facto},
which requires significant manual effort.  Consequently, the analysis
of additional data would require the availability of additional
resources.)

\begin{figure}[t]
  \centerline{\vbox to 6pc{\hbox to 10pc{}}}
  \includegraphics[scale=.4]{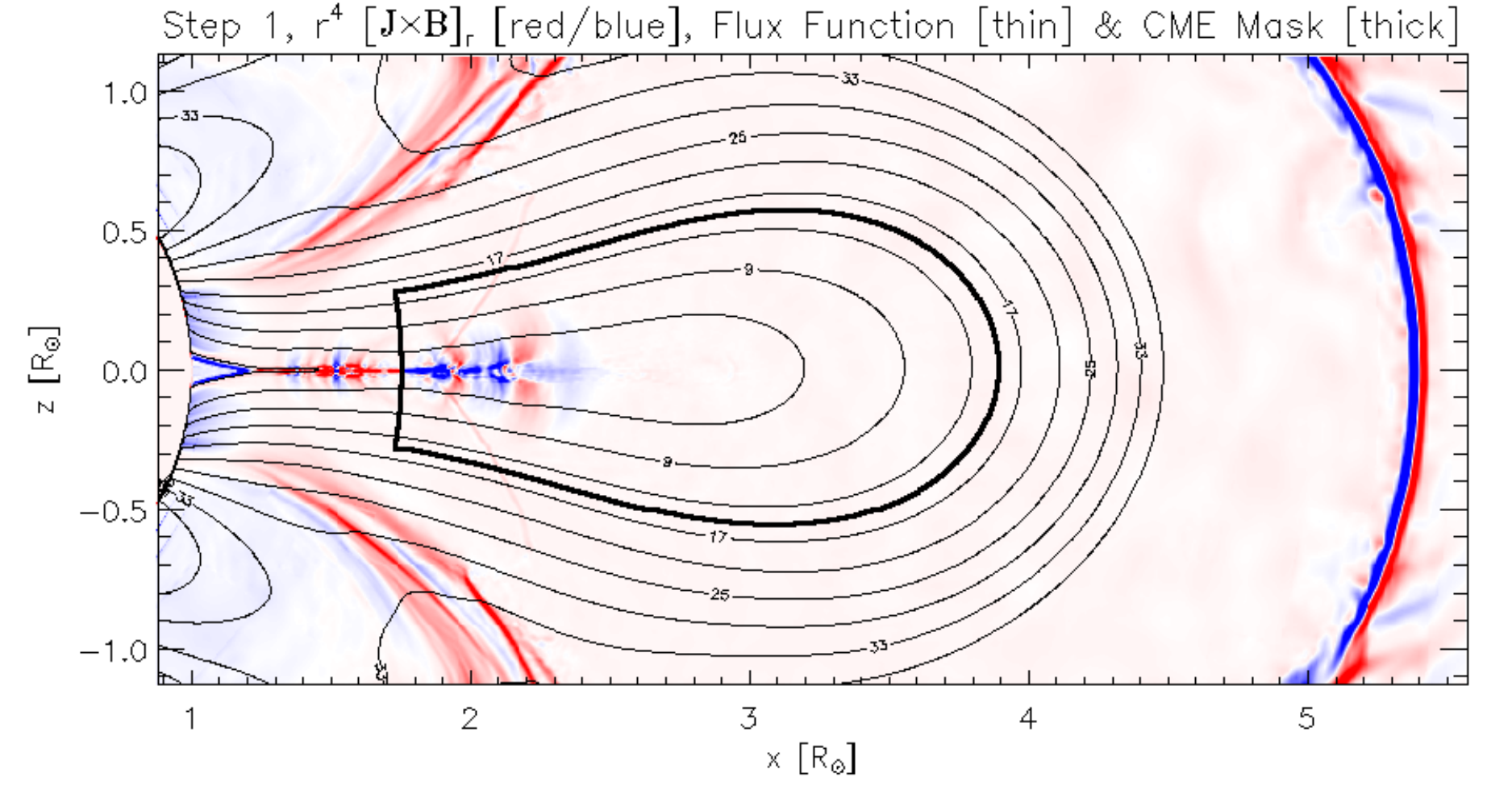}
  \includegraphics[scale=.4]{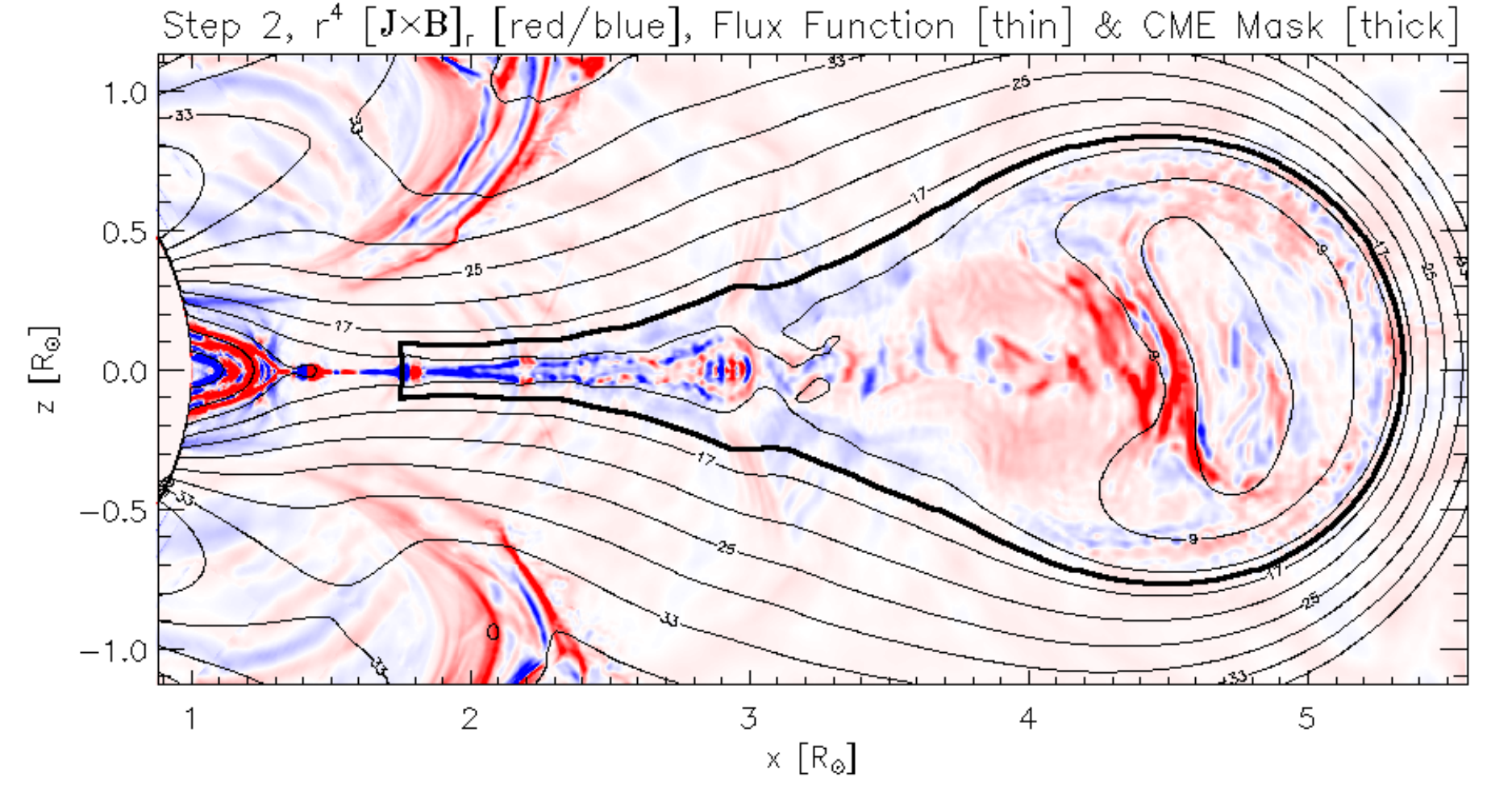}
  \caption{Red (blue) shows inward (outward) components of the radial
    Lorentz force at snapshots 1 and 2, weighted by $r^4$, so their
    spatial structure can be discerned at large $r$, where
    $\mathbf{J}$ and $\mathbf{B}$ are much weaker than near the
    model's inner boundary.  The thin black contours show the flux
    function, from which evolution of the field, including some
    effects of reconnection, can be inferred.  The thick contour shows
    a CME mask whose outer boundary is at the same flux-function
    contour in both snapshots (see text).  }
  \label{fig:LFr_steps}
\end{figure}

Figure~\ref{fig:LFr_steps} shows the inward and outward (red and blue,
respectively) components of the radial Lorentz force at each step,
both heavily saturated and weighted by $r^4$, so their spatial
structure can be discerned at large $r$, despite $\mathbf{J}$ and
$\mathbf{B}$ being much weaker there than near $r = R_\odot$. The thin black
contours show the flux function, which makes clear that a substantial
amount of flux has reconnected between these snapshots.
The thick
contour
%, at $f$ = 15,
outlines a {\em CME mask}, meant to capture the main body of the
erupting structure.  We have chosen to define this mask using a
fixed-$r$ inner boundary and an outer boundary set at the same
flux-function contour in both snapshots.
We also note the presence of several key structural features of the
global magnetic field. Post-reconnection, poloidal fields in the flare
arcade are prominent near $z$ = 0 and $x \in [1, 1.5] R_\odot$ in the
bottom panel.  The breakout current sheet is near $x = 5 R_\odot$ in
the top panel.  The separatrix current sheets, which flow between the
erupting flux system and adjacent systems and are extensions of the
breakout current sheet, run diagonally near $1.5 R_\odot < x < 2.25
R_\odot$ in both panels.

The distribution of the radial Lorentz force is richly structured,
with features in the reconnection outflow region clearly making large
contributions to the Lorentz force.  Notably, forces present in the
jet are strong enough to deform post-reconnection flux-function
contours, making them concave inward --- i.e., with a magnetic tension
that points radially inward. It is unclear whether observed CMEs
exhibit such structure.  To quantify evolution of the radial Lorentz
force as the reconnection proceeds, we can integrate the interpolated
force density over the CME mask on our Cartesian grid, weighted by $r$
(since $dV =
%(J_\theta B_\phi - J_\phi B_\theta)
dx \, dz \, r \, d\phi$, as in cylindrical coordinates).  Applying the Step-1
mask to both steps, the outward Lorentz force increases by a factor of
2.3; applying the Step-2 mask to both steps, the outward Lorentz force
increases by a factor of 1.4; and applying the Step-1 mask at Step 1
and the Step-2 mask at Step 2 (i.e., advecting the flux-function
boundary), the outward Lorentz force increases by a factor of 1.5.
%
%If flare reconnection were not ongoing between these steps, the outward
%expansion of the eruption, combined with conservation of flux, would
%weaken its magnetic field, thereby {\em decreasing} the outward
%Lorentz force.
%
%IDL> mean(mask1*bmag[*,*,0])
%    0.0075111429787516129
%IDL> mean(mask2*bmag[*,*,1])
%    0.0061639338696336697
%IDL> help,where(mask1 ne 0)
%
%IDL> 0.0061639338696336697/0.0075111429787516129
%      0.82063860
%
Expansion of the eruption, combined with conservation of flux, weakens
its magnetic field: we averaged each step's $|\mathbf{B}|$ over the
CME mask at that step, and this average decreases by about 20\%
between Steps 1 and 2.
%
%IDL> help,where(mask1 ne 0)
%<Expression>    LONG      = Array[55260]
%IDL> help,where(mask2 ne 0)
%<Expression>    LONG      = Array[91492]
%IDL> 91492./55260.
%       1.6556641
%
(The area of Step 2's mask is 5/3 the size of Step 1's.)
%
%IDL> mean(mask1*jmag[*,*,0])
%    0.0016582447778294308
%IDL> mean(mask2*jmag[*,*,1])
%    0.0022198686107063595
%IDL> 0.0022198686107063595/0.0016582447778294308
%       1.3386858
%
All else equal, a weaker $|\mathbf{B}|$ would decrease the Lorentz
force.
%, but all else is not equal:
However, the
average current density,
$|\mathbf{J}|$, in Step 2's CME mask is about 4/3 that in Step 1's.
This increase in Lorentz force can be explained both by reconnection
and the development of an ideal MHD instability, e.g., the torus
instability \citep{Kliem2006}.
%
%Clearly, reconnection plays a key role in the ejection's outward
%acceleration, consistent with the expectations of \citet{Lin2000} and
%\citet{Zhang2006}, and the findings of KAD12.

%LFr in fixed mask (1st-step):'
%================================'
%       15.952907
%       36.718774
%       21.403944
%
% 36.718774/15.952907 = 2.3016980

%LFr in fixed mask (2nd-step):
%================================
%       17.301077
%       23.808027
%       27.570467
%
% 23.808027/17.301077 = 1.3761009

%LFr in moving mask:
%================================
%       15.952907
%       23.808027
%     -0.32452481
%
%  23.808027/15.952907 = 1.4923943 
%

\begin{figure}[t]
  \centerline{\vbox to 6pc{\hbox to 10pc{}}}
  \includegraphics[scale=.8]{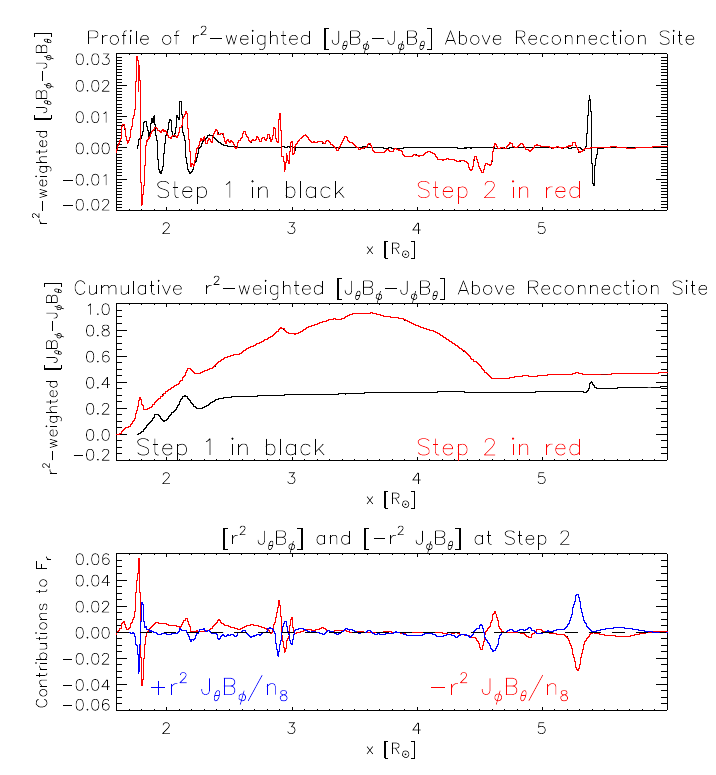}
  \caption{Top: Black (red) curves show $r^2$-weighted
    contributions to the radial Lorentz force at Step 1 (Step 2),
    averaged over a strip 100 cells (0.58 $R_\odot$) wide in $z$,
    outward in $x$ from the reconnection site at that Step.
    Middle: Cumulative sum of the curves in the top panel from the
    reconnection site to each $x$.  The excess outward force in Step 2
    compared to Step 1 arises almost entirely from the diagonal, blue
    bands in Figure \ref{fig:LFr_steps} just inside the CME mask
    around $x \in [2.5, 3.5]R_\odot$.  Flank currents (see text) flow out of the
    page in these bands (parallel to the flare-reconnection current sheet).
    Bottom: For Step 2, the blue curve shows $[r^2 J_\theta B_\phi]$ and
    the red curve shows $[-r^2 J_\phi B_\theta]$.  The latter generally
    makes a greater contribution to the outward Lorentz force. }
  \label{fig:LFr_3plots}
\end{figure}

In Figure \ref{fig:LFr_3plots}, we investigate the radial profile of
the radial component of the Lorentz force, $c F_r = (J_\theta B_\phi -
J_\phi B_\theta)$, outward from the flare reconnection site in Steps 1
and 2.
To compare contributions to $F_r$ between
small and large $r$, we weight all quantities plotted by $r^2$.
%
%weighted by $1/n_8$, where $n_8$ is the electron number density scaled
%by $10^{-8}$.  Dividing force by a measure of density implies the
%ratio yields information about the acceleration of plasma within the
%ejection.  Also, due to the simulation's density stratification, this
%rescaling enables comparing contributions to the acceleration between
%small and large $r$, since the magnetic field and density decrease by
%commensurate factors.
%
The flare reconnection site was identified in
each step by finding where the smoothed radial velocity along $z=0$
changed sign.
The values plotted in this figure are {\em
  strip-averaged}: to compute them, we averaged values of
$[J_\theta B_\phi]$ and $-[J_\phi B_\theta]$ in a 100-cell-wide
strip, encompassing $z \in [-0.29, 0.29]R_\odot$.
The highly oscillatory structure of $r^2 (J_\theta B_\phi - J_\phi
B_\theta)$ in the top panel of Figure \ref{fig:LFr_3plots}
precludes clearly identifying how the primary contributions to
the overall force arise.
To clarify which areas contribute to the CME's acceleration, in
the middle panel, at each $r$ we plot the {\em cumulative} (summed)
value of $r^2 (J_\theta B_\phi - J_\phi B_\theta)$, from the base
of the reconnection outflow to that $r$.
In this cumulative plot, a positive (negative) slope over some
interval in $r$ indicates a net positive (negative) contribution to
the CMEs outward acceleration in that interval.
Clearly,
Lorentz forces that develop in
the reconnection jet
begin making a much larger contribution to the CME's
acceleration between Steps 1 and 2.
These findings support the idea that  
reconnection plays a key role in the ejection's outward acceleration,
consistent with the expectations of \citet{Lin2000} and
\citet{Zhang2006}, and the findings of KAD12.

Crucially, by decreasing the width of the strip over which the
inverse-density-weighted radial force is averaged, we found that
essentially {\em all} of the excess contribution to the outward force
for Step 2 compared to Step 1 comes from {\em flank currents} --- the
tilted, blue bands just inside the CME mask (bottom panel of Figure
\ref{fig:LFr_steps}) for $r \in [2.5,3.5] R_\odot$.  These structures
are reminiscent of
%(and might result from)
the standing waves (slow-mode shocks) in the Petschek reconnection
model \citep{Petschek1964}, but are macroscopic regions instead of
the idealized discontinuities in that model.  We found that these
regions contain strong azimuthal currents parallel to that in the
flare reconnection sheet, i.e., out of the page, resulting from a
strong $\partial B_r/\partial \theta$ near the CME mask
boundary, in which $B_r$ exterior to the CME
%is nearly switched off
decreases
%in favor of
relative to
$B_\theta$ within the CME.  The associated component of the Lorentz
force, proportional to $B_\theta \partial B_r/\partial \theta$,
corresponds to a magnetic tension.
One expected signature of the hoop force would be substantial outward
magnetic pressure from the poloidal field over the trailing face of
the ejection, i.e., a strong $\partial B_\theta/\partial r$ (which
also contributes to $J_\phi$).  However, we see little evidence to
support this expectation.

The inward contribution to acceleration in the interval around $x \in
[3.5, 4.5]R_\odot$ in the middle panel of Figure \ref{fig:LFr_3plots}
is perhaps due to inward magnetic tension from the concave-inward
curvature of field lines near $z = 0$ in this range, as may be seen in
the bottom panel of Figure \ref{fig:LFr_steps}.  Farther from $z = 0$,
curvature in these lines reverses, implying an outward magnetic
tension that plausibly cancels much (or all) of this inward force.

Consistent with our finding about the significance of the flank
currents, the bottom panel of Figure \ref{fig:LFr_3plots} compares the
two contributions to the outward Lorentz force at Step 2, $[r^2 J_\theta
B_\phi]$ (blue) and $[- r^2 J_\phi B_\theta]$ (red).  (The sum of these
terms yields the red curve in the top panel.) There is a clear
tendency for these terms to have similar magnitude but opposite signs.
Over many intervals, however, the outward force is dominated by
$(-J_\phi B_\theta)$ --- that is, the product of the component of the
azimuthal current (the component of $\mathbf{J}$ along the PIL) and
the reconnected component of the magnetic field.
In contrast, the product of the guide magnetic field with the
current component that is redirected by reconnection from the radial
direction into the $\theta$ direction (blue vectors in the right panel of
Figure \ref{fig:shear}) is smaller at this point in the simulation.

We also remark that the volume-integrated Lorentz  force acting on a flux tube with field strength $B$ and flux $\Phi$ over the segment containing the flank currents would, like the tension force analyzed by \cite{Welsch2018}, scale as $B \Phi$. Unlike that tension force, however, the force from flank currents would only act transiently over a given tube segment, as the bend propagates along the tube.

% Units of J: R_sun*(curl B)/4pi

\section{Summary \& Prospects}
\label{sec:conclusions}

We have considered the effect of reconnection on currents and magnetic
fields in eruptive flares, to better understand (i) how reconnection
modifies these components of the Lorentz force and thus (ii) how 
flare-associated CMEs are accelerated.

We were not able to analyze enough snapshots from available ARMS simulations
to test the flux accretion model's scaling in Equation
(\ref{eqn:f_scaling}).  The necessary analysis would require computing
radial Lorentz force changes between several closely spaced simulation
steps, with each time interval involving reconnection of a known
amount of flux.  If sufficient resources --- essentially, funding ---
becomes available, such an analysis will be pursued.

We found that, in this simulation, PIL-aligned ``flank'' current structures in
the reconnection outflow make the primary contribution to the total outward
Lorentz force.  These findings accord with those reported by \citet{Jiang2021} in their analysis of another CME simulation (see bottom row of their Fig. 4).
Contrary to the prediction of the flux accretion model, we see
little evidence of additional hoop force from reconnected flux driving
further acceleration.  This conclusion, however, should not be hastily
generalized to other configurations, for two reasons, both related to
this simulation's axisymmetry: this model's reconnection jet is
probably stronger than that in real, 3D events; and this model's hoop
force is likely weaker than that in shorter-length-scale systems.
Regarding the reconnection jet: the axial invariance here implies that
flow and magnetic structures created in the reconnection process are
coherent over much longer length scales than would be expected in a
more realistic 3D geometry, within which reconnection would likely be
patchy \citep{Linton2006, Linton2009}, producing a much more
fragmented structure in the outflow.  Regarding the erupting system's
length scale: the axisymmetry here implies that the radius of
curvature of the model ejection's axis, $r_c$, is on the order of the
Sun's radius, $r_c \sim R_\odot$. This might be an apt model for the
eruption of a polar-crown filament, but does not accurately represent
the much smaller $r_c$ expected for an ejection from an active region,
for which $r_c \sim R_\odot/10$ would often be more appropriate.  Because
the strength of the hoop force increases steeply for fixed ejection
width as $r_c$ shrinks (e.g., Figure 6 in \citealt{Welsch2018}), a
much smaller $r_c$ should, all else equal, yield a much larger hoop
force.

We conclude with recommendations for future work.
In the simulation that we analyzed, the forces from the reconnection
jet have deformed the eruption's core, making its trailing edge
concave. We are unaware of observational reports of such concave
structure within CMEs; if found, this would indicate that forces in
the reconnection outflow play a substantial role in the interior
structure of CMEs.
Given the considerations in the two preceding paragraphs, we believe
that analysis of the evolution of Lorentz forces in fully 3D,
active-region-scale simulations is warranted.  We also note that,
while the association between CME speeds v$_{\rm CME}$ and
reconnection fluxes $\Phi_{\rm rec}$ for eruptive flares is relatively
well established, the flux accretion model's prediction, per the
scaling in Equation (\ref{eqn:f_scaling}), that CMEs' accelerations
(and probably their final speeds) should be correlated with coronal
magnetic field strengths in their source regions, $B_{\rm CME}$, has
not been tested.  We suggest that potential-field extrapolation might
be an easy approach to test this prediction, but note that radio
observations might enable direct
%observational
estimation of $B_{\rm CME}$.

\acknowledgements{We thank the U.S. taxpayer, whose taxes helped support the work performed here. 
 We thank
%
%C.~R. DeVore for: (1) kindly investing
%  significant effort to compute the azimuthal vector potential for some output
%  steps from one of his 2.5D ARMS breakout simulations, (2) uploading
%  these data to an accessible drive, and (3) serving as a reference
%  regarding interpretation of these data.
%  %, especially their physical units.
%  I thank
B.~J. Lynch for supplying code to read this ARMS output into IDL, and
for taking the time to write a detailed how-to for executing this
conversion.  BTW is also grateful to M. Linton and B. Kliem for their
time spent in engaging discussions of forces acting on CMEs. BTW
gratefully acknowledges support from NASA HSR-80NSSC23K0092 and NSF
SHINE AGS-2302697. CRD was supported by a grant from NASA’s H-ISFM program
to Goddard Space Flight Center.}

%\bibliographystyle{iaulike} % no .bst!!

%\bibliography{/Users/welschb/share/Latexstuff/abbrevs,%
%/Users/welschb/share/Latexstuff/short_abbrevs,%
%/Users/welschb/share/Latexstuff/full_lib,%
%/Users/welschb/share/Latexstuff/bib_mods}

% \apjl {The Astrophysical Journal Letters}
% \apj {The Astrophysical Journal}
% \Apj {The Astrophysical Journal}
% \jgr {Journal of Geophysical Research}
% \solphys {Solar Physics}
%

\section*{Discussion --- Questions \& Responses}
%\label{sec:discussion}

{\bf Question, by Dr. Bernhard Kliem:} It appears possible that the
increase of the outward hoop force from the flux accretion to the
erupting rope by flare reconnection underneath is weaker than
estimated so far, because this flux carries little current. The
currents driving the accretion of the flux flow in the slow-mode
shocks attached to the diffusion region (or to the uppermost diffusion
region if several such regions exist). The Lorentz force of these
currents accelerates the upward reconnection outflow jet, hence,
drives the accretion of flux to the erupting flux rope. The new flux
itself carries little current; it ``dipolarizes'' while joining the
rope flux. Therefore, it appears possible that the primary addition to
the upward forces consists in the momentum transfer from the upward
reconnection jet to the erupting flux rope (if the jet moves faster
than the rope). There is no doubt on the reduction of the downward
tension force of overlying flux by flare reconnection. This latter
effect is included in the treatment of the torus instability by
\citet{Kliem2006}.

{\bf Response, by Dr. Welsch:} Regarding the presence of currents in
the outflow region, the flux accretion model does indeed posit that
the post-reconnection flux {\em dipolarizes} (e.g.,
\citealt{Priest2000}) --- i.e., evolves toward a roughly dipolar
shape.  But only an exactly dipolar field is current-free; we may
sensibly speak of a compressed or stretched dipolar field, which
contains currents, but is still essentially dipolar in character
(versus unipolar or quadrupolar).  Thus, I do not use the term
``dipolarization'' to imply that significant currents are not present.
(This term appears to have originally been applied to retraction of
highly stretched magnetotail fields in geospace toward the Earth after
reconnecting; but even when fully dipolarized, these fields form a
stretched dipole.)

Clearly, however, we did not find evidence of the outward hoop force
acting in the simulation analyzed here.  However, as noted in the
concluding section of this article (above), compared to the solar
case, this simulation's axisymmetry probably makes both (1) the
reconnection jet stronger and (2) the hoop force weaker.  Hence,
determining whether the hoop force from post-reconnection magnetic
fields is dynamically important or not requires analysis of other
models with more realistic geometry.

\end{document}